\documentclass[preprint,12pt,authoryear]{elsarticle}

\usepackage{needspace}

\usepackage{algorithm}       
\usepackage{algpseudocode} 
\usepackage{amssymb}
\usepackage{amsthm}
\usepackage{amsmath}
\usepackage{booktabs} 
\usepackage{enumitem}
\usepackage{etoolbox}
\usepackage{hyperref}
\usepackage{lineno}
\usepackage{nomencl}
\usepackage{siunitx}
\usepackage{threeparttable} 

\hypersetup{
    colorlinks=true,
    linkcolor=blue,
    filecolor=magenta,      
    urlcolor=cyan,
}

\renewcommand\nomgroup[1]{\item[\bfseries\ifstrequal{#1}{S}{Subscripts}{}]}

\journal{Results in Engineering}

\begin{document}

\begin{frontmatter}

\title{Integrating a one-dimensional convolutional neural network with nonlinear model predictive control for slug flow modeling and control}

\author[inst1]{Asli Karacelik\corref{cor1}}
\ead{asli.karacelik@ntnu.no}
\cortext[cor1]{Corresponding author.}
\affiliation[inst1]{organization={Department of Mechanical and Industrial Engineering, Norwegian University of Science and Technology}, 
            city={Trondheim},
            postcode={7491}, 
            country={Norway}}

\author[inst1]{Christian Holden}
\ead{christian.holden@ntnu.no}

\begin{abstract}
Severe slugging in pipeline–riser systems degrades production efficiency and threatens equipment integrity, motivating the development of accurate yet deployable models for control. This paper presents a lightweight, data-driven modeling and control framework specifically designed for slug flow. First, we revisit a commonly used low-order mechanistic model and introduce two changes that improve physical plausibility and numerical robustness: (i) an explicit zero-outflow boundary during the starvation period at the riser top, and (ii) a reformulated expression for the liquid volume fraction at the riser top that enforces admissible bounds. Synthetic input–output trajectories generated by this model are then used to train a one-dimensional convolutional neural network (1D CNN) that parameterizes a neural ODE with an affine-in-control. Training employs an \num{8}-step free-roll objective on standardized data; a differential sensitivity check confirms an explicit control channel. The learned model is embedded in a shooting-based nonlinear model predictive control (NMPC) scheme with bounded and rate-smoothed actuation. In closed loop simulations at \(\Delta t=\SI{0.1}{\second}\), the controller drives the output to a constant setpoint within \(\sim\)\num{15}–\SI{18}{\second} while maintaining smooth, bounded inputs and non-oscillatory transients. The results indicate that a simple 1D CNN, coupled with NMPC, can capture the characteristic oscillations and regime transitions of slug flow and deliver effective regulation with modest computational cost. Although training is conducted offline to ensure reproducibility, the architecture and identification procedure are designed for online extension, providing a practical path toward adaptive slug-flow mitigation.
\end{abstract}


\end{frontmatter}

\Needspace{1\baselineskip} 
\subsection*{Subscripts}

\makenomenclature
\nomenclature{\(T\)}{Temperature}
\nomenclature{\(\rho\)}{Density}
\nomenclature{\(p\)}{Pressure}
\nomenclature{\(M\)}{Molar mass}
\nomenclature{R}{Molar gas constant}
\nomenclature[S]{\(\mathrm{g}\)}{Gas}
\nomenclature[S]{\(\mathrm{wh}\)}{Wellhead}
\nomenclature[S]{\(\mathrm{mix}\)}{Mixture}
\nomenclature{\(\alpha\)}{Volume fraction}
\nomenclature[S]{\(\mathrm{l}\)}{Liquid}
\nomenclature{\(^*\)}{Nominal condition}
\nomenclature{\(w\)}{Mass flow rate}
\nomenclature{\(A\)}{Area}
\nomenclature{\(p\)}{Pipe}
\nomenclature[S]{\(\mathrm{tot}\)}{Total}
\nomenclature{\(x\)}{Mass fraction}
\nomenclature{\(D\)}{Diameter}
\nomenclature{\(D\)}{Diameter}
\nomenclature{\(V\)}{Volume}
\nomenclature{\(m\)}{Mass}
\nomenclature{\(h\)}{Height}
\nomenclature[S]{\(\mathrm{c}\)}{Critical}
\nomenclature{\(K\)}{Correction factor}
\nomenclature{\(K_\mathrm{v}\)}{Valve coefficient}
\nomenclature{\(\Delta L\)}{Liquid-filled pipe length}
\nomenclature{\(\theta\)}{Pipe slope angle}
\nomenclature[S]{\(\mathrm{r}\)}{Riser}
\nomenclature{\(L\)}{Length}
\nomenclature[S]{\(\mathrm{rt}\)}{Riser top}
\nomenclature[S]{\(\mathrm{s}\)}{Separator}
\nomenclature[S]{\(\mathrm{out}\)}{Riser top outlet}
\nomenclature[S]{\(\mathrm{rb}\)}{Riser bottom}
\nomenclature[S]{\(\mathrm{in}\)}{Pipe inlet}
\printnomenclature

\section{Introduction}

\begin{figure*}[htbp]
 \centering
 \includegraphics[width=13cm]{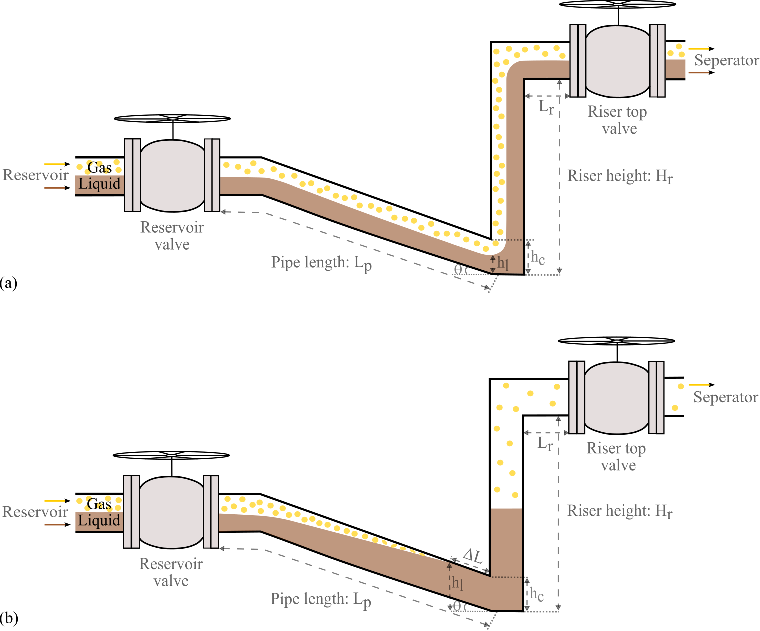}
 \caption{(a) Steady-state oil and gas production. (b) Liquid accumulation within the pipeline during slugging.}
 \label{fig:1}
\end{figure*}

Riser slugging can occur in multiphase flow systems when a downward-inclined pipeline connects to a vertical riser. A drop in reservoir pressure—and in some cases, other factors such as wax formation—can initiate this phenomenon in oil and gas production. When reservoir pressure decreases, the gas flow becomes insufficient to maintain the pressure required to push the liquid to the surface, resulting in liquid accumulation in the pipeline. Over time, this buildup forms slugs that obstruct the upward gas flow. As pressure builds behind the blockage, it eventually becomes high enough to dislodge the slugs and push them to the surface, restarting the cycle. When this flow regime leads to intermittent flow, it is called severe slugging (\autoref{fig:1}). This condition can cause significant operational issues, including corrosion, erosion, flooding, and potential system shutdowns.

Choking is the most common approach to mitigating the slugging problem. When appropriately managed, choking can effectively reduce back pressure. However, in tall-riser systems, applying choking without generating back pressure is challenging and often results in reduced production.

Another widely used method is gas lift, which helps prevent slug formation by lowering the riser's fluid density. Nevertheless, excessive gas injection can lead to gas accumulation in the riser, ultimately reducing the production rate. The extent of this reduction depends on the injection pressure, as overly high pressures can raise riser pressure and thereby lower the production rate.

Researchers have developed linear and nonlinear model predictive control (MPC and NMPC) strategies to address these issues more systematically. As the name implies, both approaches require a model to predict and control system behavior. Accurate modeling is particularly critical for NMPC to ensure smooth and stable control. Due to its linear internal model, MPC tends to be more robust than NMPC in plant-model mismatches in mildly nonlinear systems. However, when the internal model is inaccurate in NMPC, extensive tuning is often required to maintain control. In the following, we present studies that explore the application of MPC and NMPC to reduce or eliminate slugging.

Various control structures have been proposed to mitigate slug flow. \cite{Willersrud2013}, \cite{Diehl2018}, and \cite{Gerevini2018} implemented NMPC strategies using two manipulated variables. \cite{Willersrud2013} and \cite{Gerevini2018} reported a reduction in slugging, while \cite{Diehl2018} demonstrated both attenuation and complete elimination of slugging. In the control structure presented by \cite{Willersrud2013}, the manipulated variables were the production and wellhead choke valve openings, and the controlled variables were the separator inlet pressure and the line pressure at a cluster. By contrast, \cite{Gerevini2018} and \cite{Diehl2018} used lift gas mass flow rate and production choke valve opening as manipulated variables, with downhole pressure as the controlled variable.

These control strategies also contribute to production optimization. \cite{Gerevini2018} and \cite{Diehl2019} reported increased production while mitigating slugging using NMPC and MPC, respectively. Specifically, \cite{Gerevini2018} achieved a \SI{9}{\percent} increase in production by implementing NMPC with the Local Linearization on the Trajectory (LLT) algorithm. Meanwhile, \cite{Diehl2019} reported a \SI{10}{\percent} increase through a combination of regulatory and supervisory control strategies. They employed Proportional-Integral-Derivative (PID) control for regulatory tasks and used MPC as a supervisory controller to effectively suppress slugging.

Additionally, optical sensors have been employed to enhance measurement accuracy during slugging and effectively control the system using MPC. \cite{Eaton2015} integrated a fiber optic sensor upstream of the riser's touchdown zone to measure pressure. Subsequently, \cite{Moscato2024} applied an optical flow detection technique in microfluidic systems, demonstrating the broader applicability of such methods.

Recently, researchers have explored data-driven modeling approaches that are integrated with MPC and NMPC. \cite{Dalen2016} proposed subspace and parametric identification techniques — specifically Deterministic and Stochastic System Identification and Realization (DSR) and the Prediction Error Method (PEM) — and applied the resulting models within an MPC framework. \cite{Soares2022} compared NMPC responses based on different state estimation techniques, namely the Extended Kalman Filter (EKF) and a feed-forward neural network, to investigate stable and unstable operating regions in gas lift systems. Lastly, \cite{Moscato2024} constructed a linear time-invariant (LTI) data-driven model to derive a second-order transfer function.

Following the discussion on slug flow control, we now turn to machine learning-based modeling, which can provide accurate system models for MPC under the specific slugging conditions on which they were trained.

Initial studies focused on classifying and monitoring slugging in multiphase flows. For example, \cite{Ye2013} introduced an online recognition method using the support vector machine (SVM) algorithm. Later that year, \cite{Ye2013a} extended their work by incorporating statistical analysis and optimization techniques to investigate the correlation between pressure fluctuations and multiphase flow patterns. Other researchers \citep{Amr2016, Xiao2018} applied the naive Bayes algorithm to enhance classification accuracy under uncertainty.

Further studies evaluated various classifiers to identify multiphase flow. \cite{Alhashem2020} investigated multiple algorithms, including decision trees, random forests, logistic regression, SVM, and multi-layer perceptrons. \cite{Carvalho2021} compared random forests with techniques such as one-nearest neighbor, Gaussian naive Bayes, linear discriminant analysis, and quadratic discriminant analysis, concluding that random forests were the most effective classifiers. However, \cite{RoxasII2022} found that SVM outperformed random forests, feed-forward neural networks, naive Bayes, k-nearest neighbors, and gradient boosting (AdaBoost). \cite{Xu2022} also applied SVM for multiphase flow classification.

As the field progressed, the focus shifted toward forecasting multiphase flows, allowing operators to take preventive measures before slugging occurs. \cite{Andrianov2018} pioneered using long short-term memory (LSTM) networks for virtual flow metering and time series forecasting. Building on this work, \cite{Bhowmik2019} applied LSTM models to predictive maintenance scheduling, while \cite{Bangert2020} extended its application to forecasting slug flow. \cite{ShoeibiOmrani2019} compared recurrent neural networks (RNNs) with feed-forward neural networks for multiphase flow identification. More recently, \cite{Dabholkar2023} conducted a comparative study of DeepAR—an LSTM-based method—and the temporal fusion transformer (TFT), concluding that TFT outperforms DeepAR in time series forecasting.

Prior machine-learning studies on slugging primarily target detection and prevention, with LSTMs the most common choice for history-based forecasting. These approaches are typically paired with MPC or NMPC and trained offline on historical data, which constrains generalization to the operating envelopes represented in the training set. In this study, we model slug-flow dynamics with a 1D CNN embedded in NMPC to achieve active slug suppression. As outlined in Section 3.6, future work will integrate active and passive control and transition to online learning.

Physics-informed learning is increasingly used in control, either to augment mechanistic models or to regularize purely data-driven models. The choice between a physics-informed approach and a purely data-driven one should be guided by (i) the credibility and coverage of available physics, (ii) data availability and quality, (iii) real-time compute/latency budgets, and (iv) safety and assurance requirements. For highly nonlinear, complex systems, faithful first-principles models may be unavailable or too slow for real-time use; poorly specified physics can bias learning or misrepresent the state. When no reliable mechanistic model exists, a purely online, data-driven strategy can be appropriate—provided that observability/persistent excitation, measurement quality, and safety constraints are ensured.

Hybridising a credible mechanistic model with machine learning often improves sample efficiency, enforces physical plausibility (e.g., conservation and state constraints), and can extrapolate more reliably, but typically increases computational load and integration complexity. Purely data-driven models are simpler to deploy, yet performance hinges on training distribution coverage and robust sensing; they can generalise poorly under distribution shift. Online learning adapts to non-stationarity but requires reliable streaming measurements, drift detection, and safeguards against instability and catastrophic forgetting.

In safety-critical settings with a defensible physical model, physics-based or physics-informed controllers are usually preferable. Where the system evolves rapidly or physics is unreliable or non-representative, online (or other data-driven) approaches offer greater adaptability—ideally embedded within a constraint-handling, safety-aware control layer.

Because severe slug flow is difficult to model accurately and is sensitive to operating conditions, we favour online learning for deployment. Nevertheless, in this study, we train the proposed 1D CNN offline (batch) to provide a reproducible baseline and to establish feasibility. The results demonstrate that a 1D CNN can capture the characteristic oscillations and regime transitions of slug flow from input–output data, and the architecture readily extends to online (continual) updating.   

\section{Method}
The code used in this study is accessible via GitHub\footnote{\url{https://github.com/asli-karacelik/slugging-1DCNN-NMPC.git}}. We conducted this study using an Intel Core i7-10610U CPU with a base frequency of \SI{1.80}{\giga\hertz} and a maximum frequency of \SI{2.30}{\giga\hertz}.

\subsection{Mathematical model of the riser slugging}

Numerous mathematical models for slug flow have been proposed in the literature. Early work by \cite{Schmidt1980} combined experiments and modeling for pipeline–riser systems, quantifying key cycle features—such as slug buildup time, pressure fluctuations, and liquid holdup along the riser and pipeline—and showing that careful choking can mitigate severe slugging. Later, \cite{Taitel1990} proposed a mechanistic slug-flow model that uses unit-cell mass and momentum balances together with bubble-velocity relations to predict flow regimes and parameters in slug flow. Building on this line of research, \cite{Jansen1996} developed two relevant formulations: (i) a steady stability (force-balance) criterion and (ii) a simplified transient ``quasi-equilibrium'' model that estimates slug length and cycle time, extending the idea of \cite{Taitel1990} by incorporating the effects of topside choking and gas lift. 

More recently, low-order dynamic models for anti-slug control (e.g., \cite{Storkaas2003,Meglio2010,Jahanshahi2011}) represent the pipeline–riser system using a small set of coupled ordinary differential equations derived mainly from mass balances, together with simple valve and friction laws (friction is omitted in \cite{Meglio2010}). These models are intentionally simpler than detailed mechanistic formulations and have been calibrated and validated against OLGA simulations, as well as laboratory or field measurements.

In this study, we modify the model proposed by \cite{Jahanshahi2011} by reformulating the expression for the volume fraction at the top of the riser. Although they made their source code publicly available, we identified several inconsistencies between the theoretical model described in the paper and its numerical implementation.

In particular, the code imposes non-negativity constraints (i.e., zero lower bounds) on several physical variables to maintain numerical stability. Without these constraints, the simulation fails to converge or terminates due to non-physical values. While such safeguards are common and often necessary in the numerical treatment of slug-flow models — given their inherent stiffness and sensitivity — they introduce implicit assumptions that are not explicitly stated in the original publication.

Importantly, we observed that only a subset of these constraints is critical for stability in the provided implementation. Specifically, removing the lower bounds on the gas mass, liquid mass, or the liquid height \(h_\mathrm{l}\) results in solver failure or numerical breakdown.

The model constants employed in this study are summarized in \autoref{tab:1}. The mathematical formulation of the model is presented below. In the equations, the superscript \({}^*\) denotes the no-slugging (steady-state) reference operating condition.

\begin{table*}[!htbp]
    \centering
    \begin{threeparttable}
    \caption{Model constants.}
    \label{tab:1}
    \begin{tabular}{l c c c}
        \hline
        \textbf{Constant} & \textbf{Symbol} & \textbf{Value} & \textbf{Unit} \\
        \hline

        \multicolumn{4}{l}{\textbf{General Constants}} \\[2pt]
        Acceleration of gravity        & $g$                & \num{9.81}                          & \si{\meter\per\second\squared} \\
        Gas molar mass                 & $M_\mathrm{g}$     & \num{23e-3}                         & \si{\kilogram\per\mol} \\
        Liquid density                 & $\rho_\mathrm{l}$  & \num{832.2}                         & \si{\kilogram\per\meter\cubed} \\
        Liquid volume fraction\textsuperscript{*}        & $\alpha^*$         & \num{0.4}                           & -- \\
        Molar gas constant             & $R$                & \num{8.314}                         & \si{\meter\cubed\pascal\per\kelvin\per\mol} \\
        Valve coefficient              & $K_\mathrm{v}$     & \num{1}                             & -- \\

        \addlinespace[4pt]
        \multicolumn{4}{l}{\textbf{Pipe Section}} \\[2pt]
        Diameter                       & $D_\mathrm{p}$     & \num{0.12}                          & \si{\meter} \\
        Inclination angle              & $\theta$           & \num{1}                             & \si{\degree} \\
        Length                         & $L_\mathrm{p}$     & \num{4300}                          & \si{\meter} \\
        Pressure\textsuperscript{*}    & $p_\mathrm{p}^*$   & \num{70e5}                          & \si{\pascal} \\
        Temperature                    & $T_\mathrm{p}$     & \num{298.3}                         & \si{\kelvin} \\

        \addlinespace[4pt]
        \multicolumn{4}{l}{\textbf{Riser Section}} \\[2pt]
        Diameter                       & $D_\mathrm{r}$     & \num{0.1}                           & \si{\meter} \\
        Height                         & $H_\mathrm{r}$     & \num{300}                           & \si{\meter} \\
        Temperature                    & $T_\mathrm{r}$     & \num{350}                           & \si{\kelvin} \\
        Riser top length\textsuperscript{\dag} & $L_\mathrm{rt}$ & \num{100}                           & \si{\meter} \\

        \addlinespace[4pt]
        \multicolumn{4}{l}{\textbf{Well / Separator Conditions}} \\[2pt]
        Separator pressure             & $p_\mathrm{s}$     & \num{57e5}                          & \si{\pascal} \\
        Wellhead pressure              & $p_\mathrm{wh}$    & \num{70.5e5}                        & \si{\pascal} \\
        Wellhead temperature           & $T_\mathrm{wh}$    & \num{350}                           & \si{\kelvin} \\
        \hline
    \end{tabular}
    
    \vspace{4pt}
    \begin{tablenotes}
        \footnotesize
        \item \textsuperscript{*} refers to reference operating conditions where no slugging occurs.
        \item \textsuperscript{\dag} indicates the horizontal pipe section located above the riser.
        
    \end{tablenotes}
    \end{threeparttable}
\end{table*}

\subsubsection{Pipe dynamics model}
The initial conditions are defined by setting the nominal liquid and gas volume fractions in the system, $\alpha_\mathrm{l}^*$ and $\alpha_\mathrm{g}^*$, to \num{0.4} and \num{0.6}, respectively. We choose volume fractions because they provide a convenient and intuitive basis for visualizing the initial state of the system. These nominal values are assumed to be uniform and identical in both the pipe and the riser. Below, we present the corresponding calculations for the initial states.

The initial liquid mass in the pipe is 
\begin{equation}
m_{\mathrm{l,p}0} = \rho_\mathrm{l} V_\mathrm{p} \alpha_\mathrm{l}^*,
\end{equation}
where $\rho_\mathrm{l}$ denotes the liquid density, and $V_\mathrm{p}$ denotes the pipe volume.

The initial gas mass in the pipe is
\begin{equation}
m_{\mathrm{g,p}0} = \frac{p_\mathrm{wh} V_\mathrm{p} \left(1 - \alpha_\mathrm{l}^*\right) M_\mathrm{g}}{R T_\mathrm{p}},
\end{equation}
where $p_\mathrm{wh}$ is wellhead pressure, $M_\mathrm{g}$ denotes gas molar mass, $R$ represents molar gas constant, and $T_\mathrm{p}$ is the pipe temperature. In this equation, we assume the pipe pressure equals the wellhead pressure to obtain a rough estimate of the initial gas mass.

The initial liquid mass in the riser is 
\begin{equation}
m_{\mathrm{l,r}0} = \rho_\mathrm{l} V_\mathrm{r} \alpha_\mathrm{l}^*,
\end{equation}
where $V_\mathrm{r}$ is the riser volume. 

The initial gas mass in the riser is 
\begin{equation}
m_{\mathrm{g,r}0} = \frac{p_\mathrm{s} V_\mathrm{r} \left(1 - \alpha_\mathrm{l}^*\right) M_\mathrm{g}}{R T_\mathrm{r}},
\end{equation}
where $p_\mathrm{s}$ denotes the separator pressure and $T_\mathrm{r}$ denotes the riser temperature.

\cite{Jahanshahi2011} imposed constant liquid and gas mass flow rates at the pipe inlet. In this work, we calculate the flow rates based on the specified volume fractions.
To do so, we first calculate the gas density at the wellhead:

\begin{equation}
\rho_{\mathrm{g},\mathrm{wh}} = \frac{p_{\mathrm{wh}} M_{\mathrm{g}}}{R T_{\mathrm{wh}}},
\end{equation}
where $p_{\mathrm{wh}}$ is the wellhead pressure and $T_{\mathrm{wh}}$ is the wellhead temperature. 

The density of the fluid mixture at the wellhead is
\begin{equation}
\rho_{\mathrm{mix},\mathrm{wh}} = \alpha_{\mathrm{l}}^* \rho_{\mathrm{l}} + \bigl(1-\alpha_{\mathrm{l}}^*\bigr)\rho_{\mathrm{g},\mathrm{wh}}.
\end{equation}

The total mass flow rate at the wellhead is
\begin{equation}
    w_\mathrm{tot,wh}
    = K_\mathrm{v} A_\mathrm{p}
    \sqrt{\rho_\mathrm{mix,wh}\max\!\bigl[(p_\mathrm{wh}-p_\mathrm{p}^*),0\bigr]}\,,
\end{equation}
where $K_\mathrm{v}$ is the flow coefficient, $A_\mathrm{p}$ is the pipe cross-sectional area, and $p_\mathrm{p}^*$ is the nominal pipe pressure. In this equation, we set $K_\mathrm{v}=\num{1}$ and neglect valve losses. It is important to note that this valve formulation is originally derived for incompressible flow. As a result, the model requires significant tuning to accurately represent the gas–liquid system. We demonstrate this discrepancy in \cite{Karacelik2025}.

The liquid mass fraction at the pipe inlet is
\begin{equation}
x_{\mathrm{l},\mathrm{in}} = \frac{\alpha_{\mathrm{l}}^* \rho_{\mathrm{l}}}{\alpha_{\mathrm{l}}^* \rho_{\mathrm{l}} + \bigl(1-\alpha_{\mathrm{l}}^*\bigr)\rho_{\mathrm{g},\mathrm{wh}}}.
\end{equation}
Here, we assume that the liquid mass fraction is uniformly distributed throughout the pipe. Consequently, the liquid mass fraction in the pipe is equal to that at the inlet.

The liquid mass flow rate at the pipe inlet is 
\begin{equation}
w_{\mathrm{l},\mathrm{in}} = x_{\mathrm{l},\mathrm{in}} w_{\mathrm{tot},\mathrm{wh}}.
\end{equation}

The gas flow rate at the pipe inlet is 
\begin{equation}
w_{\mathrm{g},\mathrm{in}} = \bigl(1 - x_{\mathrm{l},\mathrm{in}}\bigr) w_{\mathrm{tot},\mathrm{wh}}.
\end{equation}

The nominal gas density in the pipe is
\begin{equation}
    \rho_\mathrm{g,p}^*=\frac{p_\mathrm{p}^*M_\mathrm{g}}{RT_\mathrm{p}}.
\end{equation}

The nominal liquid volume fraction in the pipe is
\begin{equation}
    \alpha_\mathrm{l,p}^*= \frac{\rho_\mathrm{g,p}^*w_\mathrm{l,in}}{\rho_\mathrm{g,p}^*w_\mathrm{l,in}+ \rho_\mathrm{l} w_\mathrm{g,in}}.
\end{equation}

The cross-sectional area of the pipe is
\begin{equation}
    A_\mathrm{p}=\frac{\pi}{4}D_\mathrm{p}^{2},
\end{equation}
where $D_\mathrm{p}$ is the pipe diameter.

The pipe volume is
\begin{equation}
    V_\mathrm{p}=A_\mathrm{p}L_\mathrm{p},
\end{equation}
where $L_\mathrm{p}$ is the pipe length. 

The nominal liquid mass in the pipe is
\begin{equation}
    {m}_\mathrm{l,p}^*= \rho_\mathrm{l}V_\mathrm{p}\alpha_\mathrm{l}^*.
\end{equation}

The nominal liquid mass in the riser is
\begin{equation}
    {m}_\mathrm{l,r}^*= \rho_\mathrm{l} V_\mathrm{r}\alpha_\mathrm{l}^*.
\end{equation}

The height of the intersection of the riser and the pipe is
\begin{equation}
    h_\mathrm{c}=\frac{D_\mathrm{p}}{\cos\theta},
\end{equation}
where $\theta$ is the pipe's inclination angle. 

The liquid level at the bottom of the riser is
\begin{equation}
    h_\mathrm{l}^*=K_\mathrm{h}h_\mathrm{c}\alpha_\mathrm{l}^*,
\end{equation}
where \( K_\mathrm{h} \) denotes the correction factor applied to tune the model. Since we did not perform any tuning in this study, we set \( K_\mathrm{h}=\num{1} \).

The liquid mass accumulation in the pipe is
\begin{equation}
    \Delta m_\mathrm{l,p}=m_\mathrm{l,p}-{m}_\mathrm{l,p}^* = \Delta L A_\mathrm{p}(1-\alpha_\mathrm{l}^*)\rho_\mathrm{l},
\end{equation}
where $m_\mathrm{l,p}$ is the liquid mass in the pipe and $\Delta L$ is the axial length of the pipe segment that is filled with liquid (single-phase liquid occupying the full cross-section).

The height of the liquid-filled pipe is
\begin{subequations}\label{eq:hl_fill}
\begin{align}
  h_\mathrm{l} &= h_\mathrm{l}^* + \Delta L \sin\theta, \label{eq:hl_fill_a}\\
  h_\mathrm{l} &= h_\mathrm{l}^* +
  \biggl(\frac{m_\mathrm{l,p}-m_\mathrm{l,p}^*}{A_\mathrm{p}(1-\alpha_\mathrm{l}^*)\rho_\mathrm{l}}\biggr)\sin\theta. \label{eq:hl_fill_b}
\end{align}
\end{subequations}

The gas volume in the pipe is
\begin{equation}
    V_\mathrm{g,p}= V_\mathrm{p}-\frac{m_\mathrm{l,p}}{\rho_\mathrm{l}}.
\end{equation}

The gas density in the pipe is
\begin{equation}
    \rho_\mathrm{g,p}=\frac{m_\mathrm{g,p}}{V_\mathrm{g,p}},
\end{equation}
where $m_\mathrm{g,p}$ is the gas mass in the pipe. 

The pressure in the pipe is
\begin{equation}
    p_\mathrm{p}=\frac{\rho_\mathrm{g,p}RT_\mathrm{p}}{M_\mathrm{g}}.
\end{equation}

\subsubsection{Riser dynamics model}

The cross-sectional area of the riser is
\begin{equation}
    A_\mathrm{r}=\frac{\pi}{4}D_\mathrm{r}^{2},
\end{equation}
where $D_\mathrm{r}$ is the riser diameter. 

The volume of the riser system is
\begin{equation}
    V_\mathrm{r}=A_\mathrm{r}(H_\mathrm{r}+L_\mathrm{rt}),
\end{equation}
where $H_\mathrm{r}$ and $L_\mathrm{rt}$ are the riser height and length of the horizontal section located above the riser.

The gas volume in the riser is
\begin{equation}
    V_\mathrm{g,r}=V_\mathrm{r}-\frac{m_\mathrm{l,r}}{\rho_\mathrm{l}},
\end{equation}
where $m_\mathrm{l,r}$ is the liquid mass in the riser. 

The gas density at the riser top is
\begin{equation}
    \rho_\mathrm{g,rt}=\frac{m_\mathrm{g,r}}{V_\mathrm{g,r}},
\end{equation}
where $m_\mathrm{g,r}$ is the gas mass in the riser. 

The pressure at the riser top is
\begin{equation}
    p_\mathrm{rt}=\frac{\rho_\mathrm{g,rt}RT_\mathrm{r}}{M_\mathrm{g}},
\end{equation}
where $T_\mathrm{r}$ is the temperature at the riser. 

The density of the mixture in the riser is 
\begin{equation}
  \rho_\mathrm{mix,r}= \frac{m_\mathrm{g,r}+m_\mathrm{l,r}}{V_\mathrm{r}}.
\end{equation}

The pressure differences at the pipe-riser intersection for the liquid and gas phases are
\begin{subequations}
\begin{align}
    \Delta p_\mathrm{l}&=p_\mathrm{p}+\rho_\mathrm{l}gh_\mathrm{l}-p_\mathrm{rt}- \rho_\mathrm{mix,r} g H_\mathrm{r}, \\
    \Delta p_\mathrm{g}&=p_\mathrm{p}-p_\mathrm{rt}- \rho_\mathrm{mix,r} g H_\mathrm{r}. 
\end{align}
\end{subequations}
where $g$ is the gravity.

The mathematical model proposed by \cite{Jahanshahi2011} does not account for the starvation period observed in severe slug flow. To capture the zero-flow interval, we introduce an additional boundary condition at the riser top for the liquid phase. Furthermore, we modify the expression for the liquid volume fraction at the riser top to ensure numerical stability and to keep the computed values within their physical limits.

Accordingly, we define the liquid volume fraction at the riser top as

\begin{subequations}\label{eq:avliqvolfrac}
\begin{alignat}{2}
\alpha_\mathrm{l,rt} &= \frac{m_\mathrm{l,r}}{V_\mathrm{r}\rho_\mathrm{l}}, &\quad& \text{if } h_\mathrm{l}\le h_\mathrm{c} \text{ and } m_\mathrm{l,r}\le m_\mathrm{l,r}^*, \label{eq:avliqvolfrac_a}\\
\alpha_\mathrm{l,rt} &= 0,                                              &     & \text{if } h_\mathrm{l}> h_\mathrm{c} \text{ and } m_\mathrm{l,r}> m_\mathrm{l,r}^*, \label{eq:avliqvolfrac_b}\\
\alpha_\mathrm{l,rt} &= 1,                                              &     & \text{if } m_\mathrm{l,r}= V_\mathrm{r}\rho_\mathrm{l}. \label{eq:avliqvolfrac_c}
\end{alignat}
\end{subequations}

The mixture density at the riser top is
\begin{equation}
  \rho_\mathrm{mix,rt}=\alpha_\mathrm{l,rt}\rho_\mathrm{l}+(1-\alpha_\mathrm{l,rt})\rho_\mathrm{g,r}.
\end{equation}

The total outflow rate at the riser top is
\begin{equation}
    w_\mathrm{tot,out}
    = K_\mathrm{v} A_\mathrm{r}
    \sqrt{2\rho_\mathrm{mix,rt}\max\!\bigl[(p_\mathrm{rt}-p_\mathrm{s}),0\bigr]}\,.
\end{equation}

The liquid mass fraction at the riser top is
\begin{equation}
    x_\mathrm{l,rt}=\frac{\alpha_\mathrm{l,rt}\rho_\mathrm{l}}{\alpha_\mathrm{l,rt}\rho_\mathrm{l}+(1-\alpha_\mathrm{l,rt})\rho_\mathrm{g,r}}.
\end{equation}

The liquid outflow rate at the riser top is
\begin{equation}
    w_\mathrm{l,out}=x_\mathrm{l,rt}w_\mathrm{tot,out}.
\end{equation}

The gas outflow rate at the riser top is
\begin{equation}
    w_\mathrm{g,out}=(1-x_\mathrm{l,rt})w_\mathrm{tot,out}.
\end{equation}

For simplicity, the gas flow area at the riser bottom is defined using a quadratic approximation, given by
\begin{subequations}\label{eq:Ag}
\begin{alignat}{2}
A_\mathrm{g} &= A_\mathrm{p}\biggl(\frac{h_\mathrm{c}-h_\mathrm{l}}{h_\mathrm{c}}\biggr)^2, &\qquad& \text{if } h_\mathrm{l}<h_\mathrm{c}, \label{eq:Ag_a}\\
A_\mathrm{g} &= 0,                                                                          &       & \text{if } h_\mathrm{l}\ge h_\mathrm{c}. \label{eq:Ag_b}
\end{alignat}
\end{subequations}

The liquid flow area at the riser bottom is
\begin{equation}
A_\mathrm{l}=A_\mathrm{p}-A_\mathrm{g}.
\end{equation}

The gas flow rate at the riser bottom is
\begin{subequations}\label{eq:wgrb}
\begin{alignat}{2}
w_\mathrm{g,rb} &= A_\mathrm{g}\sqrt{2\rho_\mathrm{g,p}\,\Delta p_\mathrm{g}}, &\qquad& \text{if } h_\mathrm{l}<h_\mathrm{c}, \label{eq:wgrb_a}\\
w_\mathrm{g,rb} &= 0,                                                          &       & \text{if } h_\mathrm{l}\ge h_\mathrm{c}. \label{eq:wgrb_b}
\end{alignat}
\end{subequations}

The liquid mass flow rate at the riser bottom is
\begin{equation}
      w_\mathrm{l,rb}=A_\mathrm{l} \sqrt{2\rho_\mathrm{l}\Delta p_\mathrm{l}}.
\end{equation}

Mass balance equations for gas and liquid phases in the pipe and riser are
\begin{subequations}\label{eq:mb}
\begin{alignat}{2}
\dot{m}_\mathrm{g,p} &= w_\mathrm{g,in} - w_\mathrm{g,rb}, &\qquad& m_\mathrm{g,p} \ge 0, \label{eq:mb_gp}\\
\dot{m}_\mathrm{l,p} &= w_\mathrm{l,in} - w_\mathrm{l,rb}, &      & m_\mathrm{l,p} \ge 0, \label{eq:mb_lp}\\
\dot{m}_\mathrm{g,r} &= w_\mathrm{g,rb} - w_\mathrm{g,out},&      & m_\mathrm{g,r} \ge 0, \label{eq:mb_gr}\\
\dot{m}_\mathrm{l,r} &= w_\mathrm{l,rb} - w_\mathrm{l,out},&      & m_\mathrm{l,r} \ge 0. \label{eq:mb_lr}
\end{alignat}
\end{subequations}

\subsection{One-dimensional convolutional neural networks}

CNNs were initially developed for two-dimensional data, such as images. Over time, they evolved into one-dimensional (1D) \citep{Kiranyaz2016} and three-dimensional (3D) \citep{Wang2023} forms to handle other types of data. 1D CNNs are commonly used for time-series prediction and sequential data analysis, whereas 3D CNNs are applied in video analysis and medical imaging, where both spatial and temporal features are important. 

In this study, we investigated the use of 1D CNNs for time-series forecasting of slug flow. Although batch learning is generally unsuitable for non-stationary or changing environments, we adopted this approach to establish a controlled baseline. Our goal is to demonstrate that 1D CNNs represent a promising candidate for future online learning applications in slug-flow modeling. 

A 1D CNN consists of an input layer, one or more convolutional layers, fully connected (dense) layers, and an output layer. In the next section, we present the mathematical formulation of forward and backward propagation in 1D CNNs.

\subsubsection{Forward and Backward Propagation in CNN Layers}

In each CNN layer, 1D forward propagation is defined as follows:
\begin{equation}
    x_k^{l} = b_k^{l} + \sum_{i=1}^{N_l-1} \operatorname{conv1D}\left( w_{ik}^{l-1}, s_i^{l-1} \right),
\end{equation}
where \( x_k^l \) denotes the input to the \( k^{\text{th}} \) neuron at layer \( l \), \( b_k^l \) represents its bias, and \( s_i^{l-1} \) corresponds to the output of the \( i^{\text{th}} \) neuron at the previous layer \( l-1 \). The weight \( w_{ik}^{l-1} \) connects the \( i^{\text{th}} \) neuron in layer \( l-1 \) to the \( k^{\text{th}} \) neuron at layer \( l \). The operation \( \operatorname{conv1D} (\cdot, \cdot) \) performs a 1D convolution without zero-padding, commonly referred to as a valid convolution. As a result, the dimension of the input array \( x_k^l \) is smaller than that of the output arrays \( s_i^{l-1} \). The intermediate output \( y_k^l \) is then obtained by applying the activation function \( f(\cdot) \) to \( x_k^l \), as:

\begin{equation}
    y_k^l = f\left( x_k^l \right) \quad \text{and} \quad s_k^l = y_k^l \downarrow ss,
\end{equation}
where \( s_k^l \) denotes the output of the \( k^{\text{th}} \) neuron at layer \( l \), and ``\( \downarrow\, \text{ss} \)'' indicates a subsampling operation with a reduction factor \(\text{ss}\).

The back-propagation algorithm operates by propagating the error in reverse from the output layer of the multilayer perceptron. Let the input and output layers be indexed by \( l = 1 \) and \( l = L \), respectively. Define \( N_L \) as the number of classes in the dataset. For a given input vector \( p \), along with its corresponding target vector \( \mathbf{t}^p \) and the output vector \( \left[ y_1^L, \dotsc, y_{N_L}^L \right]' \), the mean squared error (MSE) \( E_p \) for input \( p \) at the output layer is defined as follows:

\begin{equation}
    E_p = \text{MSE}\left( \mathbf{t}^p, \left[ y_1^L, \dotsc, y_{N_L}^L \right]' \right) = \sum_{i=1}^{N_L} \left( y_i^L - t_i^p \right)^2.
\end{equation}

To compute the derivative of \( E_p \) concerning each network parameter, the delta error \( \Delta_k^l = \frac{\partial E}{\partial x_k^l} \) must be evaluated. In particular, for updating the bias of the neuron and the weights connecting to neurons in the preceding layer, the chain rule of differentiation can be applied as follows:

\begin{equation}
    \frac{\partial E}{\partial w_{ik}^{l-1}} = \Delta_k^l y_i^{l-1} 
    \quad \text{and} \quad 
    \frac{\partial E}{\partial b_k^l} = \Delta_k^l.
\end{equation}

Thus, from the first multilayer perceptron layer to the final CNN layer, the standard scalar back-propagation is carried out as follows:

\begin{equation}
\frac{\partial E}{\partial s_k^l} = \Delta s_k^l = \sum_{i=1}^{N_{l+1}} \frac{\partial E}{\partial x_i^{l+1}} \frac{\partial x_i^{l+1}}{\partial s_k^l} = \sum_{i=1}^{N_{l+1}} \Delta_i^{l+1} w_{ki}^{l}.
\end{equation}

After completing the initial back-propagation (BP) step from layer \( l+1 \) to the current layer \( l \), the BP process can be extended to the input delta \( \Delta_k^l \) of the CNN layer \( l \). Let the zero-order upsampled map be denoted as \( \text{us}_k^l = \text{up}\left( s_k^l \right) \). The delta error is then formulated as follows:

\begin{equation}
\Delta_k^l = \frac{\partial E}{\partial y_k^l} \frac{\partial y_k^l}{\partial x_k^l} = \frac{\partial E}{\partial \text{us}_k^l} \frac{\partial \text{us}_k^l}{\partial y_k^l} f'\left( x_k^l \right) = \text{up}\left( \Delta s_k^l \right) \beta f'\left( x_k^l \right),
\end{equation}
where \( \beta = \left( \text{ss} \right)^{-1} \). Subsequently, the back-propagation (BP) of the delta error, denoted as \( \Delta s_k^l \overset{\Sigma}{\leftarrow} \Delta_i^{l+1} \), is given by:

\begin{equation}
\Delta s_k^l = \sum_{i=1}^{N_{l+1}} \operatorname{conv1Dz} \left( \Delta_i^{l+1}, \operatorname{rev} \left( w_{ki}^l \right) \right),
\end{equation}
where \( \operatorname{rev}(\cdot) \) denotes the reversal of the array, and \( \operatorname{conv1Dz}(\cdot, \cdot) \) performs full 1D convolution with zero-padding. The weight and bias sensitivities are then computed as follows:

\begin{equation} \label{eq:8}
\frac{\partial E}{\partial w_{ik}^l} = \text{conv1D}\left( s_k^l, \Delta_i^{l+1} \right) 
\quad \text{and} \quad 
\frac{\partial E}{\partial b_k^l} = \sum_{n} \Delta_k^l(n).
\end{equation}

Once weight and bias sensitivities have been computed, they are utilized to update biases and weights using the learning rate \( \varepsilon \) as follows:

\begin{equation} \label{eq:9}
w_{ik}^{l-1}(t+1) = w_{ik}^{l-1}(t) - \varepsilon \frac{\partial E}{\partial w_{ik}^{l-1}} 
\quad \text{and} \quad 
b_k^l(t+1) = b_k^l(t) - \varepsilon \frac{\partial E}{\partial b_k^l}.
\end{equation}

The forward and backward propagation through the hidden layers of the CNN are shown in. The sensitivity \( \Delta s_k^l \) corresponding to the neuron \( k^{\text{th}} \) in layer \( l \) is calculated by propagating the delta errors \( \Delta_i^{l+1} \) backward from layer \( l+1 \), as specified in \eqref{eq:8}. The propagation processes between the deepest CNN layer and the initial hidden layer of the multilayer perceptron are detailed in. Comprehensive descriptions of the BP algorithm are available in~\cite{Kiranyaz2016}.

The iterative algorithm for backpropagation applied to 1D raw signals in the training set is presented in Algorithm \ref{alg:1}.

\begin{algorithm}[H]
\caption{Back-Propagation (BP) Procedure for 1D Raw Signals} \label{alg:1}
\begin{algorithmic}[1]
\State \textbf{Initialize} network weights and biases randomly, e.g., \( \sim U(\num{-0.1}, \num{0.1}) \).
\For{each BP iteration}
    \For{each input sample \( p \) in the training set}
        \State \textbf{Forward Propagation (FP):} Propagate the input through the network to compute neuron activations \( s_i^l \), for all \( i \in [\num{1}, N_l] \) and \( l \in [\num{1}, L] \).
        \State \textbf{Backward Propagation (BP):} Calculate the delta errors \( \Delta_k^l \) by propagating the errors backward from the output layer to the input layer, for all \( k \in [\num{1}, N_l] \) and \( l \in [\num{1}, L] \).
        \State \textbf{Post-Processing (PP):} Evaluate the weight and bias sensitivities based on \eqref{eq:8}.
        \State \textbf{Update:} Adjust network weights and biases by applying the scaled sensitivities with the learning rate \( \varepsilon \), as indicated in \eqref{eq:9}.
    \EndFor
\EndFor
\end{algorithmic}
\end{algorithm}

\subsection{Data and preprocessing}
Using the proposed mathematical model, we generated univariate time-series data for \num{11} different valve openings. The dataset consists of multiple trajectories stored in a MATLAB file. Each trajectory includes \num{501} samples of input and output data, with the valve opening $u_t$ varying between \num{0.5} and \num{1.0}. Each trajectory is a two-column time series $(u_t, y_t)$, where $u_t$ denotes the manipulated input (valve opening) and $y_t$ represents the measured output (liquid outflow rate). From each trajectory, we extract overlapping windows of fixed length $W=\num{16}$.

For single-step supervision, each window contains the stacked history $[y_{t-W+1:t},\,u_{t-W+1:t}]$ and the corresponding next output $y_{t+1}$. For multi-step supervision, the same history is paired with an \num{8}-step future segment of inputs $(u_t,\dots,u_{t+7})$ and the corresponding outputs $(y_{t+1},\dots,y_{t+8})$.

We fit a z-score normalization once on the entire training corpus using all historical values of $y$ and $u$. During training and inference, histories and current inputs are normalized with these statistics.

\subsection{Output dynamics model: convolutional neural ODE with affine-in-control structure}
The output evolution is modeled via an affine-in-control with history-dependent coefficients:
\begin{equation}
\dot{y}(t) = f(h_t) + g(h_t)u(t),
\end{equation}
where $h_t$ denotes features extracted from the recent history of $(y,u)$. Concretely, a 1D CNN processes the stacked history $[y_{t-W+1:t},\,u_{t-W+1:t}]$ and produces two scalar heads at the most recent time index: a \emph{drift} term $f(h_t)$ and a \emph{gain} term $g(h_t)$. 

The discrete-time predictor applies a forward-Euler step on normalized variables:
\begin{align}
    \hat{y}_{t+1}^{(\mathrm{n})} &= y_t^{(\mathrm{n})}
    + \Delta t\big(f(h_t) + g(h_t)u_t^{(\mathrm{n})}\big),
    \qquad \Delta t = \SI{0.1}{\second},
\end{align}
which provides an explicit, instantaneous pathway for $u_t$ to influence the output increment via the learned $g(\cdot)$, while allowing the recent history to modulate both drift and gain through convolutional features.

The model architecture summarized in \autoref{tab:2} includes convolutional, max-pooling, dropout, flattening, and dense layers. It comprises \num[exponent-mode = input]{21021} trainable and \num[exponent-mode = input]{42045} optimizer parameters, resulting in \num[exponent-mode = input]{63066} total parameters.

\begin{table*}[htbp]
\centering
\caption{Model summary (Conv1D Neural-ODE with affine-in-control drift).}
\label{tab:2}
\begin{tabular}{lcc}
\toprule
\textbf{Layer (type)} & \textbf{Output shape} & \textbf{\# parameters} \\
\midrule
Input history $(x,u)$ & $(\num{16}, \num{2})$ & \num{0} \\
Conv1D ($\num{2}\!\to\!\num{64}$, $k{=}\num{5}$, pad \num{2}) & $(\num{16}, \num{64})$ & \num{704} \\
Tanh & $(\num{16}, \num{64})$ & \num{0} \\
Conv1D ($\num{64}\!\to\!\num{64}$, $k{=}\num{3}$, pad \num{1}) & $(\num{16}, \num{64})$ & \num{12352} \\
Tanh & $(\num{16}, \num{64})$ & \num{0} \\
Last-step slice ($L{:}\num{16}\!\to\!\num{1}$) & $(\num{1}, \num{64})$ & \num{0} \\
Head\,: Conv1D drift ($\num{64}\!\to\!\num{1}$, $k{=}\num{1}$) & $(\num{1}, \num{1})$ & \num{65} \\
Head\,: Conv1D gain  ($\num{64}\!\to\!\num{1}$, $k{=}\num{1}$) & $(\num{1}, \num{1})$ & \num{65} \\
Elementwise affine combine $\,(\text{drift} + \text{gain}\cdot u_t)$ & $(\num{1})$ & \num{0} \\
Concat with zero $\dot{u}$ (output $[\dot{x},\,\num{0}]$) & $(\num{2})$ & \num{0} \\
\midrule
\multicolumn{2}{r}{\textbf{Total parameters}} & \textbf{\num{13186}} \\
\multicolumn{2}{r}{Trainable / Non-trainable} & \num{13186} \;/\; \num{0} \\
\bottomrule
\end{tabular}

\vspace{0.25em}
\footnotesize\emph{Notes.} Shapes exclude the batch dimension and list (length, channels) for 1D feature maps. 
Both convolutions use a stride of \num{1}; padding preserves the temporal length. 
Parameter counts follow the standard formula \(\text{out}\times\text{in}\times k + \text{out}\) (bias enabled).
\end{table*}

\subsection{Multi-step training objective}
Training proceeds on mini-batches with a ``free-roll'' multi-step loss. For each training sample, the model is rolled forward for $L=\num{8}$ steps using the provided sequence of future inputs; at each step, the freshly predicted state is appended to the history before predicting the next step (i.e., no teacher forcing beyond the first step). The objective is the mean-squared error (MSE) between predicted and true next states, averaged over the rollout horizon and the batch. Optimization uses Adam (learning rate \num{1e-3}), weight decay \num{1e-5}, gradient-norm clipping for stability, batch size \num{1024}, and \num{150} epochs (see hyperparameters in \autoref{tab:3}). Normalization is active throughout training, allowing the network to operate on standardized variables.

\begin{table*}[htbp]
\centering
\caption{Hyperparameters for the learned dynamics model and training.}
\label{tab:3}
\begin{tabular}{ll}
\toprule
\textbf{Hyperparameter} & \textbf{Value} \\
\midrule
History window length $W$ & \num{16} \\
Sampling time $\Delta t$ & \SI{0.1}{\second} \\
Architecture (trunk) & Conv1D $\num{2}\!\to\!\num{64}$ \\
Conv kernel sizes (trunk) & $[\num{5},\,\num{3}]$ \\
Conv padding (trunk) & $[\num{2},\,\num{1}]$ \\
Conv stride (trunk) & \num{1} \\
Activations (trunk) & tanh \\
Output heads & Two Conv1D $\num{64}\!\to\!\num{1}$ ($\num{1}{\times}\num{1}$): drift and gain \\
Normalization & Z-score (fit on all training histories) \\
Loss & Multi-step MSE over \num{8}-step free roll \\
Batch size & \num{1024} \\
Epochs & \num{150} \\
Optimizer & Adam \\
Learning rate & \num{1e-3} \\
Weight decay & \num{1e-5} \\
Gradient clipping (L2) & \num{5.0} \\
Training rollout length & \num{8} steps \\
Random seed & \num{42} \\
\bottomrule
\end{tabular}

\vspace{0.25em}
\footnotesize\emph{Initialization.} PyTorch defaults are used:
Kaiming/He uniform for convolution weights; biases are initialized uniformly in
$\big[-1/\!\sqrt{\text{fan\_in}},\,1/\!\sqrt{\text{fan\_in}}\big]$. \text{fan\_in} denotes the number of input connections to a neuron.
\end{table*}

\subsection{Sensitivity sanity check}
After training, a differential sensitivity check verifies that the predicted next state $\hat{x}_{t+1}$ responds immediately to the current input $u_t$. We compute $\partial \hat{x}_{t+1}/\partial u_t$ by automatic differentiation and compare it against a symmetric finite-difference estimate. An agreement between these values provides a quick consistency check of the affine-in-control pathway.

\subsection{Model predictive control (shooting formulation)}
A receding-horizon controller is constructed atop the learned model. At each control step, a length-$N=\num{40}$ sequence of future inputs is optimized via a shooting method.

Decision variables are unconstrained parameters $\{z_k\}_{k=1}^N$ that are mapped through
\begin{align} \label{eq:uk}
    u_k &= u_\mathrm{max}\tanh(z_k), \qquad u_\mathrm{max} = \num{1}.
\end{align}
thereby enforcing the hard magnitude bound $|u_k|\le u_{\max}$.

For a candidate input sequence, the learned dynamics simulate the state trajectory forward over the horizon using the most recent history window and the forward-Euler update above. The history is updated at every step with the just-predicted state and the candidate input to preserve the correct dependence on recent behavior.

The cost functional combines: (i) average setpoint-tracking $\|x_k-r_k\|^2$ (weight $q=\num{2000}$); (ii) terminal tracking $\|x_N-r_N\|^2$ (weight $q_T=\num{5000}$); (iii) a slope-matching term penalizing $\Delta x_k-\Delta r_k$, where $\Delta x_k := x_k-x_{k-1}$ (weight $q_V=\num{50}$), to encourage gentle trajectories; (iv) input energy $\|u_k\|^2$ (weight $r=\num{1e-4}$); and (v) input-rate penalty $\|u_k-u_{k-1}\|^2$ (weight $r_{\Delta u}=\num{2e-2}$). The overall objective is the mean of the stage terms plus the weighted terminal term (see controller hyperparameters in \autoref{tab:4}).

\begin{table*}[htbp]
\centering
\caption{Hyperparameters for the receding-horizon controller.}
\label{tab:4}
\begin{tabular}{ll}
\toprule
\textbf{Hyperparameter} & \textbf{Value} \\
\midrule
Prediction horizon $N$ & \num{40} \\
Reference & Constant setpoint $r_t \equiv \num{0.5}$ \\
Input bound & $|u_k|\le \num{2.0}$ (via $u_k=\num{2}\,\tanh z_k$) \\
Stage tracking weight $q$ & \num{2000} \\
Terminal tracking weight $q_T$ & \num{5000} \\
Slope-matching weight $q_V$ & \num{50} \\
Input energy weight $r$ & \num{1e-4} \\
Input-rate weight $r_{\Delta u}$ & \num{2e-2} \\
Solver & L--BFGS (strong Wolfe) \\
Max L--BFGS iterations & \num{100} \\
Passes per control update & \num{3} \\
Applied-move low-pass $\alpha$ & \num{0.8} \\
Sampling time $\Delta t$ & \SI{0.1}{\second} \\
\bottomrule
\end{tabular}
\end{table*}

The problem is solved with L--BFGS (strong Wolfe line search). To accelerate convergence, the previous solution is shifted forward by one step to initialize the next iteration of the optimization. A small number of passes/iterations is used per control update.

Only the first input of the optimized sequence is applied (receding-horizon principle). This move is blended with the previously applied input via a first-order low-pass filter,
\begin{equation}
  u^{\mathrm{applied}}_t
  =
  (1-\alpha)\,u^{\mathrm{applied}}_{t-1} + \alpha\,u^\star_t,
  \qquad \alpha = \num{0.8},
\end{equation}
where $u^{\star}_{t}$ denotes the optimal control action at time $t$. This smoothing mitigates residual actuation chatter and improves closed-loop robustness.

\subsection{Closed-loop evaluation}
A closed-loop simulation of \SI{30}{\second} at sampling time $\Delta t=\SI{0.1}{\second}$ is executed against a constant reference $r_t \equiv \num{0.5}$ (in the physical units of $x$). At each time step, the controller optimizes a new input sequence using the current history, applies the filtered first move, and advances the learned model by one step to produce the next state. The resulting state and input trajectories are recorded and plotted for inspection.

\subsection{Software and implementation notes}
The implementation is in PyTorch. Data loading from MATLAB files is robust to format variations (with fallbacks to HDF5-based loaders). Random seeds are fixed for both NumPy and PyTorch to enhance reproducibility. The weights, horizons, and sampling times reported above match the configuration used in the provided implementation.

In brief, the pipeline (i) learns an affine-in-control, history-aware input-output model using a convolutional feature extractor; (ii) validates immediate control sensitivity; and (iii) closes the loop with a smooth, bounded, shooting-based NMPC that penalizes tracking error, input energy, and input rate, applying only a filtered first move at each receding step.

\section{Results and discussion}

The limited predictive accuracy of both mechanistic and data-driven models for severe slugging hinders real-time control, particularly under changing operating conditions. Most mitigation strategies—such as topside choking, active feedback control of a topside valve, or the use of gas lift—operate by increasing backpressure on the pipeline–riser system. Higher backpressure raises wellhead/pipeline pressure and reduces reservoir drawdown, which tends to lower the production rate. Careful control design can minimize this throughput penalty by applying just enough backpressure to suppress the instability while keeping the valve as open as possible; however, this trade-off becomes more difficult as riser height increases because the larger hydrostatic head strengthens the slugging mechanism and narrows the feasible operating window.

Previous studies have demonstrated the effectiveness of Long Short-Term Memory (LSTM) networks in capturing long-range temporal dependencies from the extensive historical data typical of slug-flow systems. In contrast, the present study shifts the focus toward pattern recognition. CNNs may offer greater advantages in this regard, particularly under real-time learning conditions, by efficiently extracting spatial–temporal features from short input windows. Moreover, the simpler architecture of CNNs compared with LSTMs makes them more suitable for online or real-time implementation in slugging analysis. 

\subsection{Mathematical model of slug flow}

\begin{figure*}[htbp]
    \centering
    \includegraphics[width=13cm]{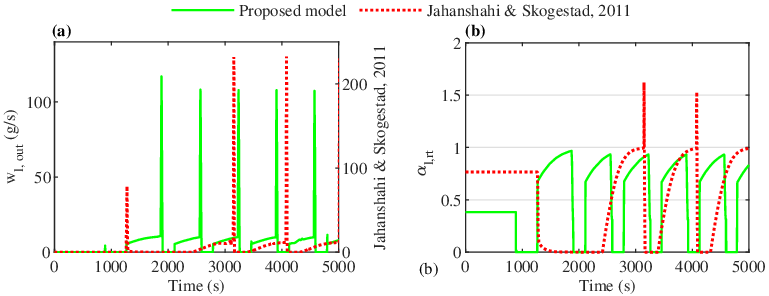}
    \caption{Comparison between the proposed slug-flow model and the reference model of \citet{Jahanshahi2011}. $w_\mathrm{l,out}$ denotes liquid outflow rate and $\alpha_\mathrm{l,rt}$ denotes liquid volume fraction at the riser top.}
    \label{fig:2}
\end{figure*}

In this study, we employ a data-driven model trained in batch (offline) mode, with the intention of extending it to online learning in future work. The training dataset is generated using the proposed mathematical model. Initially, we examined the formulation of \citet{Jahanshahi2011}; however, it shows inconsistencies with fundamental physical constraints. Specifically, the predicted liquid volume fraction at the riser top exceeds the physically admissible limit, reaching values as high as \num{1.6} (see \autoref{fig:2}). To prevent solver divergence in the presence of such nonphysical values, they introduce non-negativity bounds on selected variables to ensure numerical stability; however, these constraints were not explicitly reported in the original publication.

In the proposed model, we neglect head losses. Estimating these losses is nontrivial and typically requires calibration against laboratory or field data; therefore, they are omitted here to avoid introducing additional uncertain parameters. 

Furthermore, we modified the formulation of the liquid volume fraction at the riser top to reproduce the starvation period characteristic of severe slug flow. In contrast to \citet{Jahanshahi2011}, which models the post-blowout liquid outflow as gradual, field observations indicate a more abrupt recovery \cite{Li2022}. Accounting for this sharper transition may further increase the predicted slugging frequency in our simulations (see \autoref{fig:2}). It is also worth noting that \citet{Jahanshahi2011} calibrated their model using a correction factor to match OLGA simulations, whereas no such correction was applied in the present study. Consequently, the two models are not directly comparable.

As shown in \autoref{fig:2}, the liquid outflow rate predicted by \citet{Jahanshahi2011} is approximately twice that obtained in the present study. Modifying the liquid volume fraction at the riser top also affects the predicted outflow rate. In the experimental investigation of \citet{Li2022}, it was reported that simulation models generally overestimate the liquid outflow rate relative to laboratory measurements. In particular, their comparison of OLGA simulations with experimental data revealed that the predicted outflow rates were roughly twice the measured values. This observation suggests that our model, which yields lower outflow rates, may more accurately reflect the actual experimental behaviour—although this should be confirmed through direct validation against experimental data.

\subsection{Closed-loop tracking with learned dynamics}
\begin{figure*}[htbp]
    \centering
    \includegraphics[width=13cm]{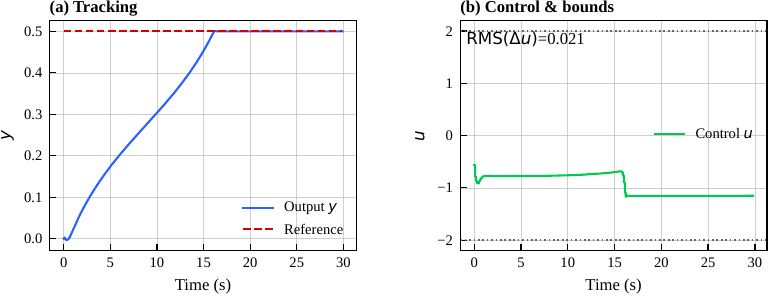}
    \caption{Closed-loop tracking and control effort using the learned Conv1D neural-ODE model and shooting-based NMPC.}
    \label{fig:3}
\end{figure*}

\autoref{fig:3} (a) illustrates that the closed-loop controller successfully drives the system state $x$ from zero to the constant setpoint $y_\mathrm{ref}=\num{0.5}$ within approximately \num{15}–\SI{18}{\second}, after which the trajectory remains steady. 
This behavior corresponds to the constant reference scenario defined in the code. 
Because the neural ODE model is designed to be \emph{affine-in-control},
\begin{equation}
    \dot{y} = f_\theta(h_t) + g_\theta(h_t) \, u_t,
\end{equation}
where $h_t$ denotes the latent history features extracted by the 1D convolutional trunk, the controller maintains an explicit and differentiable mapping from the input $u_t$ to the predicted next output. 
This property was verified through the sensitivity analysis, which confirmed that $\partial y_{t+1} / \partial u_t \neq \num{0}$, ensuring that the MPC optimizer can meaningfully adjust the control action.

The transient response in \autoref{fig:3} (a) is smooth and non-oscillatory, a direct consequence of the MPC cost structure:
\begin{equation}
    J = q \, (\text{stage}) + q_\mathrm{T} \, (\text{terminal}) + q_\mathrm{V} \, (\text{slope}) + r \, \|u\|_2^2 + r_{\Delta u}  \|\Delta u\|_2^2,
\end{equation}
where the slope and rate-penalty terms suppress high-frequency oscillations and enforce actuator smoothness. 
Thus, the monotone rise of the state is not only a property of the learned model but also of the controller design itself.

\subsection{Control effort, bounds, and rate smoothing}
\autoref{fig:3} (b) presents the control input $u$ and its saturation limits. 
In the implementation, the control signal guarantees boundedness by construction, see \eqref{eq:uk}. 

To suppress residual oscillations, the controller actively blends each new control move with the previous input, assigning \SI{80}{\percent} weight to the current optimization result and \SI{20}{\percent} to the past command. This weighted update smooths the actuator trajectory, preventing abrupt changes.
The small discontinuities in the figure (around $t=\num{10}$–\num{15}\,s) correspond to the optimizer updates, but the applied control remains within the bounds and exhibits smooth variations.

The reported $\mathrm{RMS}(\Delta \hat{u})=\num{0.021}$ confirms that the rate penalty term ($R_{\Delta u}=\num{2e-2}$) effectively limits rapid control changes. 
This is crucial for flow-assurance problems, where aggressive valve actuation can exacerbate pressure fluctuations. 
Hence, the learned-model MPC achieves accurate setpoint tracking while maintaining actuator-friendly behavior.

\subsection{Training performance of the Conv1D Neural ODE}
\begin{figure*}[htbp]
\centering
\includegraphics[width=13cm]{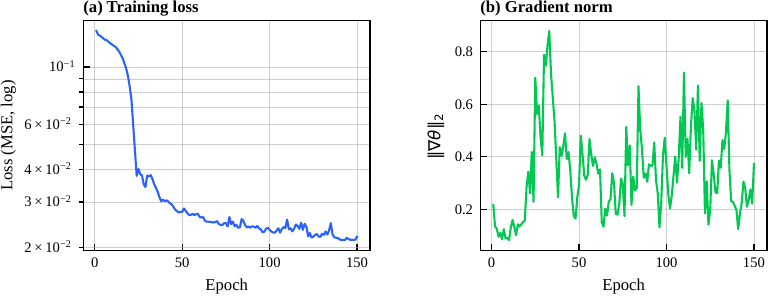}
\caption{Training performance of the Conv1D neural-ODE model. (a) Evolution of the mean-squared error loss plotted on a logarithmic scale.}
\label{fig:4}
\end{figure*}

The training evolution of the neural ODE is summarized in \autoref{fig:4}. 
Subplot (a) shows the mean-squared error loss, plotted on a logarithmic scale. 
The loss rapidly decreases during the first $\sim\num{20}$ epochs, then gradually converges to approximately $\num{2e-2}$, consistent with normalized data and mild regularization (weight decay $=\num{1e-5}$). 
Because training employs multi-step rollouts of \num{8} consecutive prediction steps, the network learns history-dependent, input-driven state transitions rather than isolated single-step mappings.

The gradient norm shown in \autoref{fig:4} (b) remains bounded throughout training. 
Transient peaks occur around epochs \num{30}–\num{60}, indicating higher-curvature regions of the loss surface, yet no signs of gradient explosion or vanishing are observed. 
The training remains stable because the gradient norm is clipped to a maximum value of \num{5.0}, thereby preventing excessively large updates during backpropagation, while the smooth tanh activations in the convolutional layers further moderate parameter changes.

\subsection{Implications for slugging control}
Taken together, the results in \autoref{fig:3}–\autoref{fig:4} demonstrate that the proposed \emph{Conv1D Neural-ODE} with an \emph{affine-in-control} structure, coupled with a bounded and rate-smoothed MPC, can accurately reproduce and control the nonlinear input–output dynamics of the slugging system. 
The framework offers three key benefits:
\begin{enumerate}[label=(\roman*)]
    \item it can be trained purely from measured input–output trajectories $(u_t, y_t)$ without requiring explicit physical parameters;
    \item it provides a differentiable control channel, enabling direct use in optimization-based control;
    \item it ensures stable and smooth control actions under realistic actuation limits.
\end{enumerate}

\subsection{Directions for future research}

In future work, we aim to implement online learning with real-time control to mitigate slug flow actively. Two control strategies will be applied: active and passive control. Pressure data will be continuously monitored at the wellhead and recorded during slug-flow events.

When slugging occurs, the active control strategy will first be engaged. The choke valve position will be dynamically adjusted until the wellhead pressure reaches the threshold required to suppress slug formation. This approach relies on real-time feedback and is expected to provide rapid stabilization.

Once active control has reduced the oscillations, the passive control strategy will be initiated. Gas will be injected into the reservoir, and the corresponding pressure response at the wellhead will be monitored. As the pressure gradually increases, the valve opening will be incrementally enlarged and eventually fully opened. At this stage, the gas injection rate will be adjusted proportionally to the measured gas outflow rate, thereby maintaining the wellhead pressure passively and stably. It should also be noted that increasing gas injection alters the gas–liquid ratio in the system, which in turn may modify the slugging behaviour. 

This approach not only mitigates slugging but also increases production rates, albeit at a higher operational cost. Moreover, if carbon dioxide is injected into the reservoir, it may contribute to sustainable oil and gas production. However, carbon dioxide injection also introduces additional operational challenges that warrant further investigation.

One of the main challenges in this approach is its strong dependence on data quality. Sensor drift represents a major source of error when working with measurement data. However, such drift may be detected by analyzing correlations between sensors installed at different positions, assuming that the sensors do not drift in the same manner. Using sensors from different manufacturers at separate locations might further help to reduce the risk of identical drift patterns and improve overall reliability. Consequently, addressing sensor drift as a system-level phenomenon rather than as isolated sensor errors can lead to more effective detection and compensation strategies.

\section{Conclusion}

We presented an end-to-end framework for modeling and control of severe slug flow that combines a compact Conv1D neural-ODE model with a shooting-based NMPC. On the modeling side, we modified the formulation of liquid volume fraction at the riser top to (i) enforce a starvation-period zero-outflow boundary and (ii) keep the variables within admissible bounds, improving physical realism and numerical robustness. Using trajectories generated by this model, we identified an affine-in-control neural ODE via multi-step (free-roll) supervision and verified, through a differential sensitivity check, that the current input directly influences the predicted increment. On the control side, we embedded the learned dynamics in an NMPC with hard magnitude limits, explicit move-rate penalties, and a first-order move filter. Closed-loop simulations ($\Delta t=\SI{0.1}{\second}$) showed rapid, monotone convergence to the setpoint (\(\sim\)\num{15}–\SI{18}{\second}) with smooth, bounded actuation and no oscillatory overshoot, demonstrating that a small convolutional architecture suffices to reproduce and regulate slug-flow dynamics under the studied conditions.

The approach is intentionally simple and computationally economical, which makes it a promising candidate for real-time deployment. At the same time, several limitations define clear next steps. (i) The identification data are generated by a simplified mechanistic model (e.g., head losses neglected), so systematic validation against laboratory experiments and field measurements is needed. (ii) Training was performed offline; implementing online/continual learning with safeguards against instability and catastrophic forgetting is a priority for operation under changing conditions. (iii) Extending the framework to multivariable sensing and actuation (e.g., combining topside choking with gas-lift) and incorporating measurement-noise handling and sensor-drift diagnostics will improve robustness. Addressing these points will help translate the demonstrated feasibility into a practical, adaptive slug-mitigation solution for production systems.

\bibliographystyle{elsarticle-harv} 
\bibliography{article3_ref}

\end{document}